\documentclass[runningheads]{llncs}
\usepackage{graphicx,times,amsmath,amssymb,cite,subfigure,url}
\usepackage[lined,ruled,commentsnumbered]{algorithm2e}
 \allowdisplaybreaks
\newcommand{\hytt}[1]{\texttt{\hyphenchar\font=\defaulthyphenchar #1}}
\begin{document}

\mainmatter
\title{EEG-Based Driver Drowsiness Estimation Using Convolutional Neural Networks}
\titlerunning{EEG-Based Driver Drowsiness Estimation Using Convolutional Neural Networks}

\author{Yuqi Cui and Dongrui Wu}
\authorrunning{EEG-Based Driver Drowsiness Estimation Using Convolutional Neural Networks}

\institute{School of Automation\\ Huazhong University of Science and Technology, Wuhan, Hubei, China\\
\email{yuqicui@hust.edu.cn}, \email{drwu@hust.edu.cn}}
\maketitle

\begin{abstract}
Deep learning, including convolutional neural networks (CNNs), has started finding applications in brain-computer interfaces (BCIs). However, so far most such approaches focused on BCI classification problems. This paper extends EEGNet, a 3-layer CNN model for BCI classification, to BCI regression, and also utilizes a novel spectral meta-learner for regression (SMLR) approach to aggregate multiple EEGNets for improved performance. Our model uses the power spectral density (PSD) of EEG signals as the input. Compared with raw EEG inputs, the PSD inputs can reduce the computational cost significantly, yet achieve much better regression performance. Experiments on driver drowsiness estimation from EEG signals demonstrate the outstanding performance of our approach.

\end{abstract}

\section{Introduction}

Drowsy driving is one of the most important causes of traffic accidents, following only to alcohol, speeding, and inattention \cite{Sagberg2004}. As a result, it is very important to monitor the driver's drowsiness level and take actions accordingly. There have been many different approaches \cite{Abbood2014,Chacon-Murguia2015,Sahayadhas2012,Kang2013} for doing so, which can be roughly categorized into two groups:
\begin{enumerate}
\item \emph{Contactless detection approaches}, which do not require the driver to physically wear any sensors. Their main advantage is the convenience to use. Contactless detection approaches can be further classified into two categories:
    \begin{enumerate}
    \item \emph{Computer vision based detection approaches}, which can be applied to either the driver or the vehicle.

        When applied to the driver, a typical practice is to place some cameras behind the windshield, which capture the driver's head in realtime. From the video we can compute the eye blink frequency \cite{Dinges1998,Ji2004},  the percentage of eye closure (PERCLOS) \cite{Sommer2010,Dinges1998a}, the eye movement \cite{Edwards2007,Eriksson1997}, the head pose \cite{Dinges1998,Murphy-Chutorian2010}, etc, which are indicators of drowsiness. The main drawback of these approaches is that they can be easily affected by the lighting condition.

        When applied to the vehicle, usually some cameras are used to capture the relative position of the vehicle in the lane. From lane departure events we can estimate the driver drowsiness \cite{Edwards2007,Chacon-Murguia2015,Sahayadhas2012}. The main drawback of this approach is that it can also be easily affected by lighting and weather, and it may not work when the lane markers are unclear or missing.

    \item \emph{Driver-vehicle interaction based detection approaches}, which use various sensors to measure the driving patterns, e.g., speeding, tailgating, abrupt braking, inappropriate steering wheel adjustments, etc \cite{Krajewski2009,Sahayadhas2012}, to infer if the driver is drowsy.
    \end{enumerate}
\item \emph{Contact sensor based detection approaches}, which require the driver to physically wear some sensors to measure his/her physiological signals, e.g., electroencephalogram (EEG) \cite{drwuTFS2017,Michail2008,drwuaBCI2015,drwuEBMAL2016}, electrocardiography \cite{Jahn2005,Michail2008}, electromyography \cite{Akin2008,Hu2009}, respiration \cite{Sharma2015,Tayibnapis2016}, galvanic skin response \cite{Edwards2007,Boon-Leng2015}, etc. Theoretically, physiological signals are more accurate and reliable drowsiness indicators, as they originate directly from the human body. Their main disadvantages include: 1) the driver's body movements may introduce artifacts and noise to the physiological signals, and hence reduce the detection accuracy; and, 2) the driver may feel uncomfortable to wear such body sensors.
\end{enumerate}

This paper focuses on the contact sensor based detection approaches. More specifically, we consider EEG-based driver drowsiness detection. The main reason is that EEG signals, which directly measure the brain state, have the potential to predict the drowsiness before it reaches a dangerous level. Hence, compared with other approaches, there is ample time to alert the driver to avoid accidents.

There has been research on using deep learning \cite{Hajinoroozi2015a,Hajinoroozi2016} for driver drowsiness classification. This paper considers regression instead of classification. It makes the following three contributions:
\begin{enumerate}
\item It extends EEGNet \cite{Lawhern2016}, a convolutional neural network (CNN) originally designed for classification problems in brain-computer interface (BCI), to regression problems.
\item It uses spectral meta-learner for regression (SMLR) \cite{drwuSMLR2016}, an unsupervised ensemble regression approach, to aggregate multiple EEGNet regression models for improved performance.
\item Instead of using raw EEG signals as the input to EEGNet, it uses their power spectral density (PSD) at certain frequencies as the input, which significantly saves the computational cost, and also improves the regression performance.
\end{enumerate}

The remainder of this paper is organized as follows: Section~\ref{sect:method} introduces our proposed EEGNet-PSD-SMLR approach. Section~\ref{sect:results} presents the details of a drowsy driving experiment in a virtual reality (VR) environment, and the performance comparison of EEGNet-PSD-SMLR with several other approaches. Finally, Section~\ref{sect:conclusions} draws conclusions and points out a future research direction.

\section{The EEGNet-PSD-SMLR Model} \label{sect:method}

This section introduces our proposed EEGNet-PSD-SMLR model for driver drowsiness estimation.

\subsection{EEGNet for Regression} \label{sect:EEGNet}

The CNN regression model used in this paper is modified from the EEGNet classification model \cite{Lawhern2016}, which has demonstrated outstanding performance in four different BCI applications, i.e., P300 visual-evoked potential, error-related negativity, movement-related cortical potential, and the sensory motor rhythm.

Denote an EEG epoch as $\mathbf{x}\in \mathbb{R}^{C\times T}$, where $C$ is the number of channels and $T$ is the number of time samples (or features) per channel. The EEGNet classification and regression architectures are given in Table~\ref{tab:EEGNet}, where $N$ is the number of classes in classification. Observe that the two architectures are identical for the first three layers; the only difference occurs at the fourth layer. The EEGNet classification architecture uses softmax regression for classification, whereas the EEGNet regression architecture uses a dense layer followed by an activation layer for regression. We have tested different activation functions (ReLU, sigmoid, tanh, and linear), and found linear activation gave the best results. So, linear activation was adopted in this paper.

\begin{table}[h]\centering \setlength{\tabcolsep}{1mm}
\caption{EEGNet architectures for classification and regression.}   \label{tab:EEGNet}
\begin{tabular}{|c|cccc|}\hline
 Layer  & Input Size & Operation & Output Size & Number of Parameters  \\\hline
 1	&	$C\times T$ 		 	&	16$\times $Conv1D(C,1) 	&	$16\times 1\times T$		& 	$16C+16$\\
 	& 	$16\times 1\times T$ 	& 	BatchNorm			&	$16\times 1\times T$ 	&      32\\
	&	$16\times 1\times T$	&	Reshape				&	$1\times 16\times T$		&	\\
	&	$1\times 16\times T$	&	Dropout(0.25)			&	$1\times 16\times T$		&	\\ \hline
2	&	$1\times 16\times T$	&	4$\times $Conv2D(2,32)	&	$4\times 16\times T$		&	$4\times 2\times 32+ 4=260$ \\
	&	$4\times 16\times T$	&	BatchNorm			&	$4\times 16\times T$		&	8\\
	&	$4\times 16\times T$	&	Maxpool2d(2,4)			&	$4\times 8\times T/4$	&	\\
	&	$4\times 8\times T/4$	&	Dropout(0.25)			&	$4\times 8\times T/4$	&	\\\hline
3	&	$4\times 8\times T/4$	&	4$\times $Conv2D(8,4)	&	$4\times 8\times T/4$	&	$4\times 4\times 8\times 4+ 4=516$ \\
	&	$4\times 8\times T/4$	&	BatchNorm			&	$4\times 8\times T/4$	&	$8$\\
	&	$4\times 8\times T/4$	&	Maxpool2d(2,4)		&	$4\times 4\times T/16$	&	\\
	&	$4\times 4\times T/16$	&	Dropout(0.25)		&	$4\times 4\times T/16$	&	\\\hline
4 (Class.)	&	$4\times 4\times T/16$	&	Softmax Regression				&	$N$					&	$TN+N$	\\\hline
4 (Regr.)	&	$4\times4\times T/16$	&	Dense	&	$1$		&	$T$ or $T+1$\\
&	$1$					&	Activation				&	$1$		&	$1$	\\\hline
Total	&				&	Classification		&				&	$16C+N(T+1)+840$ \\
&						&	Regression		&				&	$16C+T+841$ \\\hline
\end{tabular}
\end{table}

\subsection{SMLR for EEGNet Regression Model Aggregation} \label{sect:SMLR}

It's well-known that neural network models can be easily trapped at local minima. Since the EEGNet regression model is compact and can be trained quickly, we can use ensemble learning to increase its robustness. More specifically, we train 10 different EEGNet regression models by bootstrapping, and then use SMLR \cite{drwuSMLR2016} to aggregate them.

Consider a regression problem with a continuous value input space $\mathcal{X}$ and a continuous value output space $\mathcal{Y}$. Assume there are $n$ unlabeled samples, $\{\mathbf{x}_j\}_{j=1}^n$, with unknown true outputs $\{y_j\}_{j=1}^n$, and $m$ base regression models, $\{f_i\}_{i=1}^m$. The $i$th regression model's prediction for $\mathbf{x}_j$ is $f_i(\mathbf{x}_j)$. The goal of SMLR is to accurately estimate $y_j$ by optimally combining $\{f_i(\mathbf{x}_j)\}_{i=1}^m$. As shown in Algorithm~\ref{alg:SMLR}, SMLR consists of two steps: 1) estimate the accuracy of each base regression model; 2) select and combine the strong base regression models.

\begin{algorithm}[h] 
\KwIn{$n$ unlabeled samples, $\{\mathbf{x}_j\}_{j=1}^n$\;
\hspace*{8mm} $m$ base regression models, $\{f_i\}_{i=1}^m$.}
\KwOut{The $n$ estimated outputs, $\{f(\mathbf{x}_j)\}_{j=1}^n$.}
Apply each $f_i$ to $\{\mathbf{x}_j\}_{j=1}^n$ to obtain the estimates $\{f_i(\mathbf{x}_j)\}_{j=1}^n$ and assemble them into a vector $\mathbf{f}_i(\mathbf{x})$\;
Compute the covariance matrix $Q\in \mathbb{R}^{m\times m}$ of $\{\mathbf{f}_i(\mathbf{x})\}_{i=1}^m$\;
Compute the first leading eigenvector, $\boldsymbol{\mu}_0$, of $Q$\;
Perform $k$-means clustering ($k=3$) on the absolute values of the elements of $\boldsymbol{\mu}_0$\;
Identify $S$, the subset of the strong regression models, as those belong to the cluster with the maximum centroid\;
\textbf{Return} $f(\mathbf{x}_j)=\frac{\sum_{i\in S} \mu_{0,i} f_i(\mathbf{x}_j)}{\sum_{i\in S} \mu_{0,i}}$, \quad $j=1,...,n$.
\caption{The SMLR algorithm \cite{drwuSMLR2016}.} \label{alg:SMLR}
\end{algorithm}

\section{Experiment and Results} \label{sect:results}

\subsection{Dataset}

The experiment setup used in this paper was identical to that in \cite{drwuaBCI2015,drwuSMLR2016}. Sixteen healthy subjects with normal or corrected-to-normal vision were recruited to participant in a sustained-attention driving experiment \cite{Chuang2012,Chuang2014}, which consisted of a real vehicle mounted on a motion platform with six degrees of freedom immersed in a 360-degree VR scene. Each subject performed the experiment for about 60-90 minutes in the afternoon when the circadian rhythm of sleepiness reached its peak. To induce drowsiness during driving, the VR scene simulated monotonous driving at 100 km/h on a straight and empty highway. During the experiment, random lane-departure events were introduced every 5-10 seconds, and participants were instructed to steer the vehicle to compensate for them immediately. Their response time was recorded and later converted to a drowsiness index (see the next subsection), as research has shown that it has strong correlation with fatigue \cite{Ji2004}. Participants' scalp EEG signals were recorded using a 500Hz 32-channel Neuroscan system (30-channel EEGs plus 2-channel earlobes).

\subsection{Preprocessing} \label{sect:prep}

The 16 subjects had different lengths of experiment, because the disturbances were presented randomly every 5-10 seconds. Data from one subject was not recorded correctly, so we used only 15 subjects. To ensure a fair comparison, we used the first 3,600 seconds data for each subject.

We defined a function \cite{drwuaBCI2015,drwuSMLR2016} to map the response time $\tau$ to a drowsiness index $y\in[0, 1]$:
\begin{align}
y=\max\left\{0,\,\frac{1-e^{-(\tau-\tau_0)}}{1+e^{-(\tau-\tau_0)}}\right\} \label{eq:y}
\end{align}
$\tau_0=1$ was used in this paper, as in \cite{drwuaBCI2015,drwuSMLR2016}. The drowsiness indices were then smoothed using a 90-second square moving-average window to reduce variations. This does not reduce the sensitivity of the drowsiness index because previous research showed that the cycle lengths of drowsiness fluctuations are longer than four minutes \cite{Makeig1993}.

We used EEGLAB \cite{Delorme2004} for EEG signal preprocessing. A 1-50 Hz band-pass filter was applied to remove high-frequency muscle artifacts, line-noise contamination and direct current drift. Next the EEG data were downsampled from 500 Hz to 250 Hz and re-referenced to averaged earlobes.

We tried to predict the drowsiness index for each subject every 3 seconds. All 30 EEG channels were used in feature extraction. We epoched 30-second EEG signals right before each sample point, computed the power spectral density (PSD) in the theta and alpha bands (4-12 Hz) for each channel using Welch's method \cite{Welch1967}, and converted them into dBs. Each channel had 67 such PSD points at different frequencies. Some channels may have dBs significantly larger than others, which degraded the regression performance. So we removed channels which had at least one dB larger than 20, and normalized the dBs of all remaining channels to mean zero and standard deviation one. Assume the number of remaining channels is $C'$ (usually $C'$ is about 30). Then, the input matrix to our EEGNet regression model has dimensionality $C'\times 67$.

\subsection{Algorithms}

We used data from 14 subjects to build a regression model for the 15th subject, simulating the scenario that we already collected data from 14 subjects and need to use their data to help estimate the drowsiness level for a new driver. We repeated this process 15 times so that each subject had a chance to be the ``new" driver.

We compared the performance of the following five algorithms:
\begin{enumerate}
\item \emph{Ridge regression based on principal component features} (\texttt{RR}), which is the baseline. This method was first used in \cite{drwuaBCI2015}. It combined data from all existing 14 subjects and extracted average PSDs in the theta band as features. Similar to the case in Section~\ref{sect:prep}, some channels may have extremely large average PSDs, which were removed (using a 20 dB threshold) for better regression performance.  We then normalized the dBs of each remaining channel to mean zero and standard deviation one, and extracted a few (usually around 10) leading principal components, which accounted for 95\% of the variance. The projections of the dBs onto these principal components were then used as our features. At last we built a ridge regression model for the 15th subject.

\item \emph{RR based on principal component features and SMLR} (\texttt{RR-SMLR}). This is the method proposed in \cite{drwuSMLR2016}. We built 14 RR models, each one using only one source subject's data as the training dataset. Feature extraction was the same as in \texttt{RR}. After obtaining 14 models trained on different datasets, we used SMLR to aggregate them for the target subject.

\item \emph{EEGNet regression model using band-passed EEG inputs} (\texttt{EEGNet}), which used the EEGNet regression architecture described in Section\ref{sect:EEGNet}. EEG signals, after 1-50 Hz band-pass filtering, were used as input. So, the input dimensionality was $30\times 7500$ (the second dimensionality was 7500 because we used 30-second EEG signals for estimation, and the sampling rate was 250 Hz).

\item \emph{EEGNet regression model using the PSD features} (\texttt{EEGNet-PSD}). The EEGNet regression architecture was identical to the one in \texttt{EEGNet}, but the $C'\times 67$ PSD features described in Section~\ref{sect:prep} were used as its input.

\item \texttt{EEGNet-PSD} \emph{with SMLR} (\texttt{EEGNet-PSD-SMLR}), which was the above \hytt{EEGNet-PSD} model combined with SMLR ensemble learning, as described in Section~\ref{sect:SMLR}.
\end{enumerate}
Each algorithm was repeated 10 times so that statistical meaningful results can be obtained. The performance measures were the root mean square error (RMSE) and the correlation coefficient (CC), as in \cite{drwuaBCI2015,drwuSMLR2016}.

\subsection{Results and Discussions}

\begin{figure}[h]\centering
\subfigure[]{\includegraphics[width=0.8\textwidth,clip]{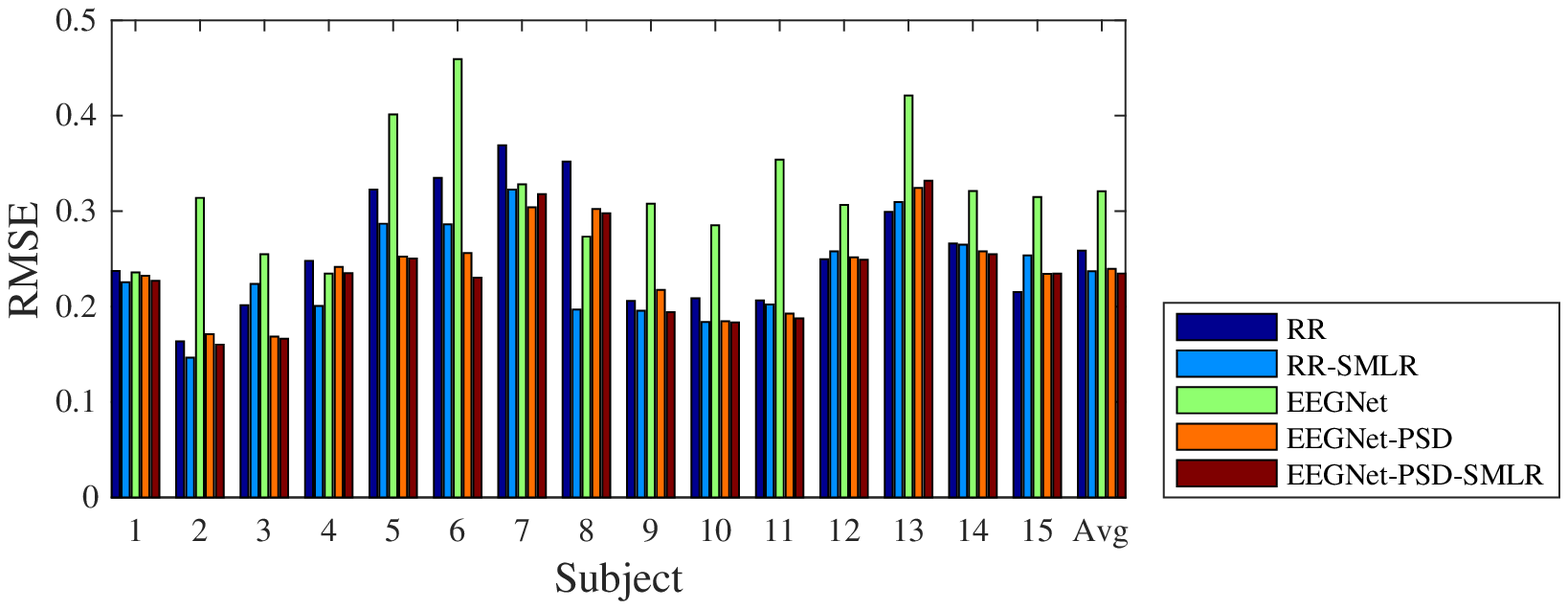}\label{fig:RMSE}}
\subfigure[]{\includegraphics[width=0.8\textwidth,clip]{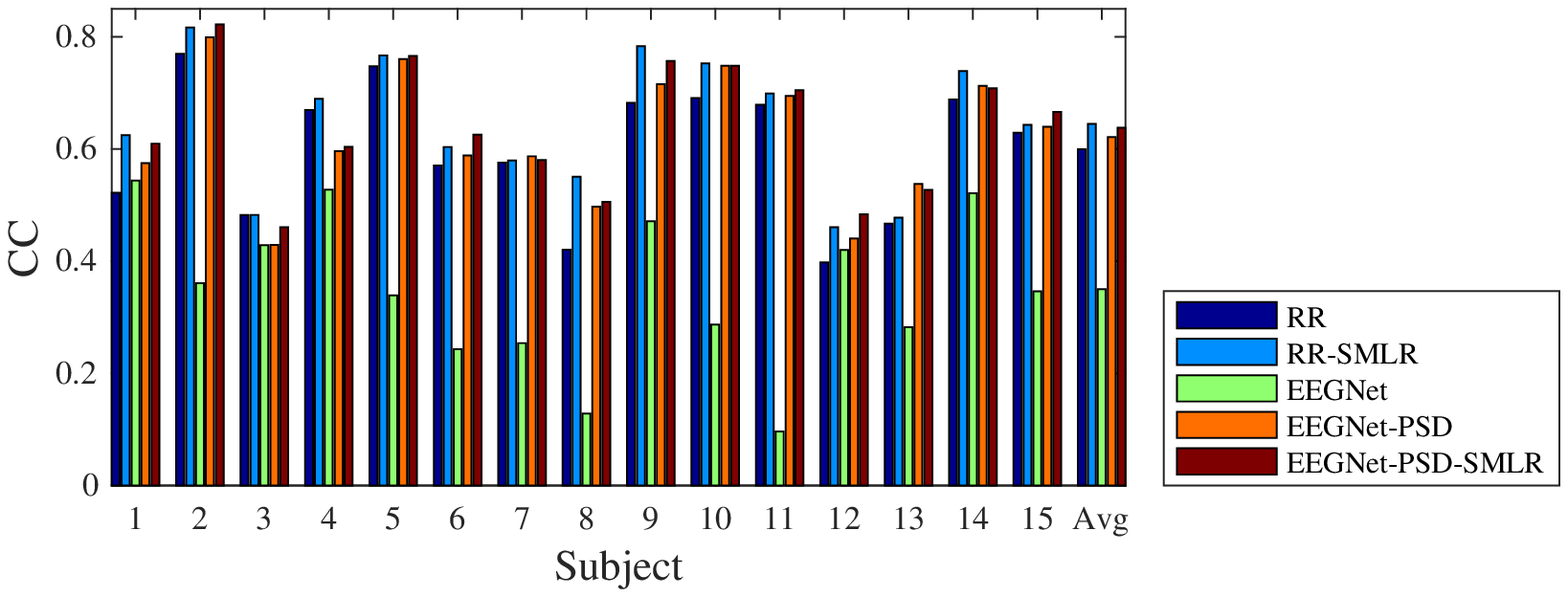} \label{fig:CC}}
\caption{(a) RMSEs and (b) CCs of the five approaches on the 15 subjects. The last group in each subfigure shows the average performance across the 15 subjects. } \label{fig:perf}
\end{figure}

The experimental results are shown in Fig.~\ref{fig:perf} and  Table~\ref{tab:AveResult}. Observe that:
\begin{enumerate}
\item \texttt{EEGNet}, which used band-passed EEG signals as the input, had the worst RMSE and CC for most subjects and also on average. This is because the input feature had very large dimensionality ($T=7500$ in Table~\ref{tab:EEGNet}), so there were  about 8820 parameters in this model. On the contrary, there were only $1200\times14=12800$ training samples, which may not be enough to fully optimize these parameters.

\item \texttt{EEGNet-PSD}, which had about 67 PSD points in each channel, achieved better RMSE and CC than both \texttt{RR} and \texttt{EEGNet} for most subjects. This demonstrates that the PSD features are better than the band-passed EEG temporal features. Because of the much smaller dimensionality, training time of \texttt{EEGNet-PSD} was also reduced significantly compared with \texttt{EEGNet}.

\item \texttt{EEGNet-PSD-SMLR}, which is an ensemble of multiple \texttt{EEGNet-PSD} aggregated by the SMLR, achieved comparable performance with \texttt{RR-SMLR}, which was our best approach on this driving dataset. On average its RMSE was $1.99\%$ smaller than \texttt{EEGNet-PSD}, and its CC was  $2.65\%$ larger than \texttt{EEGNet-PSD}. This suggests that SMLR can indeed improve the learning performance.
\end{enumerate}

\begin{table}[h]\centering \setlength{\tabcolsep}{1.5mm}
\caption{Average performances of the five algorithms on the 15 subjects.}   \label{tab:AveResult}
\begin{tabular}{c|ccccc}\hline
&	\texttt{RR}&	\texttt{RR-SMLR}&	\texttt{EEGNet}&\texttt{EEGNet-PSD}&	\texttt{EEGNet-PSD-SMLR} \\\hline
RMSE &  0.2587   & 0.2371    &0.3208    &0.2394    &$\mathbf{0.2347}$\\
CC &0.5994    &$\mathbf{0.6446}$   &0.3499    &0.6215    &0.6379\\\hline
\end{tabular}
\end{table}

We also performed a two-way Analysis of Variance (ANOVA) for the five algorithms to check if the RMSE and CC differences among them were statistically significant, by setting the subjects as a random effect. The results are shown in Table~\ref{tab:ANOVA}, which shows that there were statistically significant differences (at 5\% level) for both RMSEs and CCs.

\begin{table}[h] \centering \setlength{\tabcolsep}{2mm}
\caption{$p$-values of two-way ANOVA tests for the five algorithms.}   \label{tab:ANOVA}
\begin{tabular}{l|cc}   \hline
&RMSE & CC  \\ \hline
   $p$ & $\mathbf{<.0001}$ & $\mathbf{<.0001}$ \\ \hline
\end{tabular}
\end{table}

Then, non-parametric multiple comparison tests based on Dunn's procedure \cite{Dunn1961,Dunn1964} were used to determine if the difference between any pair of algorithms was statistically significant, with a $p$-value correction using the False Discovery Rate method \cite{Benjamini1995}. The $p$-values are shown in Table~\ref{tab:Dunn1}, where the statistically significant ones are marked in bold. Observe that the RMSE differences and the CC differences between \texttt{EEGNet-PSD-SMLR} and \texttt{RR}/\texttt{EEGNet} were statistically significant, but the differences between \texttt{EEGNet-PSD-SMLR} and \texttt{EEGNet-PSD}/\texttt{RR-SMLR} were not.

\begin{table}[h]\centering
\caption{$p$-values of non-parametric multiple comparisons for the five algorithms.}   \label{tab:Dunn1}
\begin{tabular}{c|r|cccc}\hline
&	      &	\texttt{RR}		&\texttt{RR-SMLR} &	\texttt{EEGNet}	&	\texttt{EEGNet-PSD}		\\\hline
&\texttt{RR-SMLR} & \textbf{.0040} & & &\\
&\texttt{EEGNet}			 &	\textbf{.0000}		&	\textbf{.0000}	&	&			\\
RMSE&\texttt{EEGNet-PSD}		 &	\textbf{.0087}	&	.3757 & \textbf{.0000}	&			\\
&\texttt{EEGNet-PSD-SMLR} &	\textbf{.0007}	&	.3239 & \textbf{.0000}	&	.2416		\\\hline
& \texttt{RR-SMLR} & \textbf{.0015} &&&\\
&\texttt{EEGNet}			 &	\textbf{.0000}		&	\textbf{.0000}	&	&			\\
CC&\texttt{EEGNet-PSD}		 &	.0731	& .0767 &	\textbf{.0000}	&		\\
&\texttt{EEGNet-PSD-SMLR} &	\textbf{.0055}	&.3226 &	\textbf{.0000}	&	.1550		\\\hline
\end{tabular}
\end{table}

\section{Conclusions} \label{sect:conclusions}

This paper focused on the much under-studied regression problems in BCI, particularly, driver drowsiness estimation from EEGs. It has extended EEGNet, a 3-layer CNN model for BCI classification, to BCI regression, and also utilized SMLR to aggregate multiple EEGNets for improved performance. Another novelty of our model is that it uses the PSD of EEG signals as the input, instead of raw EEG signals. In this way it can reduce the computational cost significantly, yet achieve much better regression performance. Experiments showed that  EEGNet-PSD-SMLR achieved comparable performance with our best regression model proposed recently.

Recently Riemannian geometry features have demonstrated outstanding performance in several BCI classification applications \cite{Congedo2013,Barachant2013}. Our latest research \cite{drwuRG2017} has also showed that Riemannian geometry features can outperform the traditional powerband features in an EEG-based BCI regression problem. Our future research will investigate Riemannian geometry features in the EEGNet and SMLR framework.

\end{document}